\documentclass[prl,aps,amssymb,showpacs,twocolumn]{revtex4}
\usepackage{amsmath}
\usepackage{amssymb}
\usepackage{amsthm}
\usepackage{amsfonts}
\usepackage{enumerate}
\usepackage{latexsym}
\usepackage[dvips]{graphicx}
\usepackage{epsfig}
\newcommand{\beq}{\begin{equation}}
\newcommand{\eneq}{\end{equation}}

\input{epsf}

\begin{document}

 \tolerance 10000

%\draft

\title{Orbitronics: the Intrinsic Orbital Hall Effect in p-Doped Silicon}

\author {B. Andrei Bernevig, Taylor L. Hughes, Shou-Cheng Zhang}

\affiliation{Department of Physics, Stanford University, Stanford,
CA 94305}
\begin{abstract}
%\vspace*{-1.0truecm}
\begin{center}

\parbox{14cm}{The spin Hall effect depends crucially on the
intrinsic spin-orbit coupling of the energy band. Because of the
smaller spin-orbit coupling in silicon, the spin Hall effect is
expected to be much reduced. We show that the electric field in
p-doped silicon can induce a dissipationless orbital current in a
fashion reminiscent of the spin Hall effect. The vertex correction
due to impurity scattering vanishes and the effect is therefore
robust against disorder. The orbital Hall effect can lead to the
accumulation of local orbital momentum at the edge of the sample,
and can be detected by the Kerr effect.}
\end{center}
\end{abstract}

\maketitle

Spin manipulation in semiconductors has seen remarkable
theoretical and experimental interest in recent years with the
advent of spin-electronics, and with the realization that strong
spin-orbit coupling in certain materials can influence the
transport of carriers in so-called spintronics devices
\cite{wolf2001}. Recently, a new way to manipulate spin has been
proposed, where the spin current is created by an electric field
through the intrinsic spin-orbit coupling in the semiconductor
bands\cite{murakami2003,sinova2003}. The direction of the spin
polarization and the current flow direction are mutually
perpendicular and perpendicular to the electric field. The spin
Hall effect has been recently observed experimentally by
\cite{kato2004c,wunderlich2004}.

While of possible great application in semiconductors with large
spin-orbit coupling such as GaAs and InSb, the effect is expected
to be smaller in the most used semiconductor of the electronics
industry: silicon. Indeed, the small spin-orbit coupling in
silicon, as measured by the energy of the split-off band relative
to the top of the valence band, $\sim 44\ meV$, makes the
spin-Hall effect effect small at room temperature. Recently, Yao
and Fang\cite{yao2004} computed the intrinsic spin Hall effect
from first principle, for a variety of materials including
silicon.

Given the dominance of silicon in semiconductor industry, it is
important to find a similar dissipationless transport process
which does not rely on the spin-orbit coupling. In this paper, we
investigate the possibility of replacing the spin degree of
freedom by the orbital degree of freedom, and call the associated
field of study orbitronics. The valence band of Si consists
largely of three p-orbitals. The three orbital degrees of freedom
transform as a (pseudo-) spin one quantity under rotation, are odd
under time reversal, and couple to the crystal momentum of the
hole. We show that p-doped Si under the influence of an electric
field develops an intrinsic orbital current of the $p$-band. The
polarization of the $p$-orbitals, the direction of flow, and the
direction of the electric field are mutually perpendicular. The
transport equation is similar in form to the spin-Hall
equation\cite{murakami2003}: \beq
j^{i}_{j}=\sigma_{I}\epsilon_{ijk}E_{k}. \eneq \noindent Here
$j^{i}_{j}$ stands for the orbital current flowing along the $j$
direction, where the local orbitals are polarized along the $i$
direction. For an electric field on the $y$-axis, we expect an
orbital current flowing in the positive $x$ direction to be
polarized in the $+z = p_x + i p_y$ direction while the orbital
current flowing in the negative $x$ direction is polarized in the
$-z = p_x - i p_y$ direction. Like the spin current, the orbital
current is also even under time reversal, and the above response
equation is dissipationless.

As a semiconductor with diamond structure the valence band of Si
contains 3 $p$-orbitals where the holes reside
\cite{luttinger1956}. While in most semiconductors the intrinsic
spin$-1/2$ of the holes couples with the spin$-1$ $p$-orbitals to
create the light and heavy hole bands as well as the split-off
band, in silicon this coupling is small and its energy scale is
easily overtaken by disorder or thermal fluctuations. We therefore
neglect it. The diamond lattice symmetry therefore requires that
the form of the Hamiltonian near the zone center be
\cite{luttinger1955,overhauser1956}
\begin{equation}
H = A k^2 - (A-B) \sum_{i=1}^3 k_i^2 I_i^2 - 2C \sum_{i\ne
j=1}^{3} \frac{1}{2}\{k_i, k_j\} \frac{1}{2}\{ I_i, I_j \}
\end{equation}
\noindent where the $I_i$ are the orbital angular momentum
matrices:
\begin{eqnarray*}
    I_x = \frac{1}{\sqrt{2}}  \left( \begin{array}{ccc} 0 & 1 & 0
\\ 1 & 0 & 1 \\ 0 & 1 & 0 \end{array} \right)
I_y = \frac{1}{\sqrt{2}}  \left( \begin{array}{ccc} 0 & -i & 0 \\
i & 0 & -i \\ 0 & i & 0 \end{array} \right)
I_z
=\left(\begin{array}{ccc} 1 & 0 & 0 \\ 0 & 0 & 0 \\ 0 & 0 &-1
\end{array} \right),
\end{eqnarray*}
\noindent
 and $A,B,C$ are material constants. An essential feature
of the above Hamiltonian is the coupling between the local orbital
moment $I_i$ and the momentum $k_i$. In analogy with the
spin-orbit coupling we call this orbital-orbit coupling. Within
the spherical approximation $A-B = C$ and the Hamiltonian becomes
\begin{equation}
H = A k^2 - r (\vec{k} \cdot \vec{I})^2, \label{sphericalH}
\end{equation}
\noindent where we have defined $r\equiv A-B$ to simplify
notation. This form is identical to the spherically symmetric
Luttinger Hamiltonian for the light and heavy hole bands, but as a
fundamental physical (and, as we shall see, mathematical)
difference, the matrix $I$ is not spin-$3/2$ $4\times 4$ matrix
but spin-$1$ p-orbital $3\times 3$ matrices. For simplicity, we
now work with the spherically symmetric Hamiltonian.

\begin{figure}[ht]
\includegraphics[width=5cm]{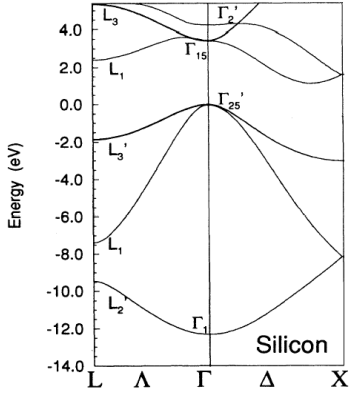}
\caption{Si energy bands from \cite{grosso1995}. The effective
Hamiltonian in Eq.(\ref{sphericalH}) describes the $\Gamma_{25'}$
bands and is valid close to the $\Gamma$ point.}
\label{bandfigure}
\end{figure}

Good quantum numbers for this Hamiltonian are the helicity
$\lambda = \vec{k} \cdot \vec{I} /k$ and the total angular
momentum $\vec{J} = \vec{x} \times \vec{p} + \vec{I}$ which is a
sum of the usual motion angular momentum plus the localized
orbital momentum. The energy bands contain of two degenerate bands
of helicity $\lambda = \pm 1$ as well as a third band of helicity
$\lambda=0$
\begin{equation}   \epsilon_{\pm
1}(k)=Ak^2 \;\;\; \epsilon_0(k) = (A-r)k^2 .
\end{equation} \noindent
Introducing five symmetric, traceless matrices\cite{murakami2004d}
$\xi_a^{ij}$, where $a=1,2,3,4,5$, $i,j=1,2,3$,
$\xi_a^{ij}=\xi_a^{ji}$ and $\xi_a^{ii}=0$, the second term in Eq.
(\ref{sphericalH}) can be expanded as:
\begin{equation}
H(k) = \epsilon(k) + r(d_a \Gamma^a)
\end{equation}
where
\begin{eqnarray}
& \epsilon(k) = \frac{k^2}{3}(3A-2r)\ ,\ d_a(k)=\xi_a^{ij} k_i k_j\ , \ \Gamma_a=\xi_a^{ij} I_i I_j, \nonumber \\
& d_1 = -\sqrt{2}k_y k_z, \;\;\; d_2= -\sqrt{2}k_x k_z , \;\;\;
d_3 = -\sqrt{2}k_x k_y \nonumber \\ & d_4 =
-\frac{1}{\sqrt{2}}(k_{x}^2-k_{y}^2),\;\;\; d_5 =-
\frac{1}{\sqrt{6}}(2k_{z}^2- k_{x}^2 - k_{y}^2).
\end{eqnarray}
Following Ref. \cite{murakami2004d}, one can similarly define the
so called conserved spin \emph{i.e.} the spin operator projected
onto the eigenstate bands of the model. Using the projection
operators onto the helicity bands in terms of the $\Gamma^a$:
\begin{eqnarray} & P_{\lambda^2 = 0}
= (1 - (\hat{k}\cdot I)^2)=\frac{1}{3}+\frac{1}{k^2}d_a \Gamma^a
\nonumber \\ &
 P_{\lambda^2
= 1} = (\hat{k}\cdot I)^2=\frac{2}{3} - \frac{1}{k^2} d_a \Gamma^a.
\end{eqnarray} (with the usual projection operator properties $P_0^2 = P_0$,
$P_{\pm 1}^2 = P_{\pm 1}$, $P_0 P_{\pm 1} = 0$) , the Hamiltonian
can be written as $H =\epsilon_{\pm 1}(k) P_{\pm 1}(k) +
\epsilon_0(k) P_{0}(k)$. The conserved local orbital moment
operator then commutes with the Hamiltonian:
\begin{equation}
I_i^{cons} = P_0 I_i P_0 + P_1 I_i P_1, \;\;\; [H, I_i^{cons}]=0;
\;\;\; i=1,2,3.
\end{equation}
\noindent The conserved orbital moment formalism physically
implies that we consider the system in its adiabatic state, where
changes to the equilibrium state, such as applied fields, etc.,
are slow enough as to maintain the system in its energy
eigenstates. It is also the case that the local orbital moment
operator $I_i$ has no projection onto the zero-helicity band, i.e.
$P_0 I_i P_0 =0$ for $i=1,2,3$. Hence, the projected motion of
local orbital moments is equivalent to the projection onto the
degenerate helicity $\lambda = \pm 1$ bands.

We now consider the effect of a uniform electric field $\vec{E}$
in our system. The application of an electric field introduces an
extra-potential $V(x)= e \vec{E} \cdot \vec{x}$. This changes the
equation of motion by acting on the particle momentum with the
obvious result of accelerating the particles in the momentum
direction. Since however, the momentum is coupled to the local
orbital moment, the electric field will influence its motion and
orientation, in a similar way as the spin-Hall effect. In
particular, a non-zero local orbital current appears which
selectively polarizes moving electrons into certain p-orbitals.

One can define two orbital currents, though one is perhaps more
appropriate. First we have the conventional orbital-current given
by $ J^{i}_{j}=\frac{1}{2} \left\{\frac{\partial H}{\partial
k_j},I^i \right\}$ with the brackets denoting an anticommutator.
This current is not conserved by the Hamiltonian dynamics and thus
a more appropriate current to consider is that given by the motion
of the conserved orbital moment $ J^{i}_{j(c)}= \frac{1}{2}
\left\{\frac{\partial H}{\partial k_j},P_0 I^i P_0 +P_{ 1} I^i
P_{1}\right\} = \frac{1}{2}\left\{\frac{\partial H}{\partial k_j},
P_{ 1} I^i P_{1}\right\}$. This conserved current is the current
of the orbital moment in the helicity $\pm 1$ band. We calculate
the DC-response of both of these currents using the Kubo formula.

The Green's function for the above Hamiltonian is matrix-valued
and reads :
 \beq
G(E,k)= (E-(\epsilon(k)+(A-B)d_a\Gamma^a))^{-1} \eneq \noindent
which involves inverting a $3 \times 3$ matrix. The lack of a
Clifford-algebra property for the $3 \times 3$ $\Gamma^a$ matrices
makes the solution of this problem much harder than in the similar
spin-Hall effect formalism \cite{murakami2004d}. However, after
some algebra and the use of the identity $P_0 P_1 \equiv 0$, the
Green's function can be written as:
\begin{equation}
G(E,k)=\frac{3(rk^2-3E(k))-9rd_a\Gamma^a}{(-3E(k)+2rk^2)(3E(k)+rk^2)},
\end{equation}
with $E(k)\equiv E-\epsilon(k)$. We compute the response function
of the spin current to an electric field. From the Kubo formula we
get:
 \beq
Q^{l}_{ij}(i\nu_m)=\frac{1}{V\beta}\sum_{k,n} Tr
(J^{l}_{i}G(i(\omega_n+\nu_m),k)J_j G(i\omega_n,k))\eneq with
Matsubara frequencies $\nu_m=2\pi
m/\beta,\omega_n=(2n+1)\pi/\beta$, and charge current operator
$J_j=\partial H/\partial k^j.$ After performing the summation over
the Matsubara frequencies, and after tracing over all the
matrices, the complicated sum is reduced to a simpler form. For
example the $Q^1_{23}$ component for the conserved current is
given by: \beq Q^1_{23}(i\nu_m)=- \frac{1}{V}\sum_{k}
\frac{(n_{F}(\epsilon_{\pm 1})-n_{F}(\epsilon_0))r^2 k_x^2
\nu_m}{2((i \nu_m)^2-r^2k^4)}.\eneq In this equation $n_{F}$ is
the Fermi-Dirac distribution function. The next step is to
consider the zero-frequency limit of the orbital conductivity
which is obtained from the response function as $
\sigma^l_{ij}=\lim_{i\nu_m\rightarrow
0}\frac{Q^l_{ij}(i\nu_m)}{\nu_m}.$ After the momentum integration,
we obtain a beautiful tensor structure: \beq
\sigma^i_{jk}=\epsilon_{ijk}\sigma_I.\eneq \noindent It is
especially suggestive that this tensor structure is identical to
the one in the spin Hall effect \cite{murakami2003} although the
gauge (matrix) structure of the Hamiltonian is fundamentally
different. The conductivity $\sigma_I$ gives:
\beq\sigma_I=\int^{k^F_{\pm 1}}_{k^F_0}\frac{d^3
k}{(2\pi)^3}\frac{k_{x}^2}{2k^4} = \frac{1}{12\pi^2}(k^F_{\pm
1}-k^F_0).\eneq $k^F_{\pm 1, 0}$ are the Fermi momenta of the two
bands. An identical picture emerges if we consider the response of
the non-conserved orbital current to an electric field, the only
difference being in the value of the constant $\sigma_I$ which in
the non-conserved case is $\sigma_I=
\frac{1}{12\pi^2}(\frac{5}{3}+4\frac{A}{A-B})(k^F_{\pm 1}-k^F_0)$.
Estimates for the orbital and charge conductivities for silicon at
a given carrier density are given in Table I. These calculations
assume the mobility of holes in silicon is $450 \ cm^2/V\cdot
s$\cite{levinstein1996}.

In the computation above we have made $3$ approximations. We now
discuss their validity. The first approximation is the neglect of
the Si spin-orbit (SO) coupling. As commented at the beginning of
the paper, the Si SO coupling is small (44 meV) and is likely to
get suppressed by the energy scale of disorder and temperature.
The energy scale at room temperature is of the same order of
magnitude as the Si SO coupling $\sim 26\ meV$, while the
difference of energies between $\epsilon_{\pm 1} - \epsilon_0$
bands at the Fermi energy is much larger \emph{e.g.} $\Delta E\sim
120\ meV$ at $n=10^{20}/cm^3$ (refer to Table I for values at
other densities). Hence the orbital Hall effect proposed here is
much larger than the spin Hall effect that Si might have due to
its small SO coupling. Moreover, our approximation gets better as
the temperature is increased.

Another caveat in this calculation is our neglect of anisotropy
when considering the spherical Luttinger Hamiltonian as a model
for silicon. We have assumed that in the Luttinger Hamiltonian
$(A-B)\approx C$ for silicon though this is not necessarily true.
When $((A-B)-C)/A$ is large we cannot form the rotationally
invariant $(\vec{k}\cdot\vec{I})^2$ term and the problem becomes
more mathematically challenging. However, the anisotropy only
slightly modifies the numerical factors given in the formulas for
$\sigma_I$.

Now we turn to the effect of impurities. In the context of the
spin Hall effect, analytical calculations have shown that the spin
Hall effect in the Rashba model\cite{sinova2003} is cancelled by
the vertex corrections due to impurity scattering\cite{inoue2004}.
On the other hand, the vertex correction vanishes
identically\cite{murakami2004b} for the spin Hall effect in the
Luttinger model describing the holes\cite{murakami2003}. This
result rests on the fact that the current vertex is odd under
parity, where the Hamiltonian is even under parity. A similar
argument is valid here: the orbital current operator is odd under
parity while the Hamiltonian is even. For this reason, the vertex
correction due to impurity scattering vanishes for the orbital
Hall effect discussed here and the effect should be robust.

Consider now an electric field parallel to the $y$-axis. In this
case, per the transport equation $J^i_j =\epsilon_{ijy} E_y$ we
have an orbital current flowing in the $j=x$ direction with
orbital local moment polarized in the $i=z$ direction. Since there
is no net charge current in the $xz$ plane, what is happening at a
microscopic level is that there are an equal number of holes
flowing in the $\pm x$ direction. However, the holes flowing in
the $+x$ direction tend to populate more the $p_x + ip_y$ local
orbitals so as to give a net $+z$ polarization, while the holes
flowing in the $-x$ direction tend to occupy predominantly $p_x -
i p_y$ local orbitals so as to give the net $-z$ polarization. At
one of the boundaries of the sample there will be a net
accumulation of $p_x + i p_y$ occupied orbitals while at the
opposite boundary the holes will tend to occupy predominantly $p_x
- i p_y$ orbitals.

We now turn to the experimental detection of our effect. There
have been several recent experiments that are setup to detect spin
currents via the associated spin accumulation at the boundary
\cite{wunderlich2004,kato2004c} and these provide us with a basis
for detecting the intrinsic orbital current in silicon. Due to the
fact that Si is an indirect-gap material with low efficiency for
light emission, an LED-type experiment like \cite{wunderlich2004},
where the polarization of the emitted light gives information
about the orbital where the emitting electron resides, is not
experimentally viable. However, Kerr and Faraday rotation
measurements are insensitive to the Si indirect gap and can be
used to probe orbital polarization.

\begin{table}

\begin{tabular}{c c c c c c}
\hline
 n($cm^{-3}$) & $\sigma_I$ ($\frac{1}{\Omega\cdot cm}$) &$\sigma_c$($\frac{1}{\Omega\cdot cm}$) &  $\ell$($\mu m$) &$\Delta E$(meV)&$\rho_I\ (\frac{\mu_B}{cm^{2}})$ \\
$10^{21}$  & 20.7 & 72000&3.9&540&$2\times 10^{15}$\\
$10^{20}$  & 9.63 & 7200&1.8&120&$9\times 10^{14}$\\
$10^{19}$  & 4.47 & 720&0.85&24&$4\times 10^{14}$\\
$10^{18}$  & 2.07 & 72&0.39&5.4&$2\times 10^{14}$\\
$10^{17}$  & 0.963 & 7.2&0.18&1.2&$9\times 10^{13}$\\
$10^{16}$  & 0.447 & 0.72&0.085&0.25&$4\times 10^{13}$\\
\hline
\end{tabular}
\caption{Parameter values are given as a function of the density
$n$. We have presented the orbital conductivity $\sigma_I$, the
charge conductivity $\sigma_c$, the spin diffusion length $\ell$,
the energy difference between the two hole bands at the Fermi
energy $\Delta E$, and the orbital polarization density $\rho_I.$}
\end{table}

\begin{table}
\begin{tabular}{c c c c}
\hline
A ($eV/m^2$)& B ($eV/m^2$) & C ($eV/m^2$) & A-B ($eV/m^2$)\\
$0.951$ & $1.51$ & $-2.06$
& $-0.558$\\
 \hline
\end{tabular}
\caption{Specific material parameters for silicon calculated using
\cite{luttinger1956,lawaetz1971}.}
\end{table}

As seen from the detection of the spin-Hall effect in GaAs the
time-resolved Kerr (Faraday) microscope is an effective
experimental apparatus for spintronics. A very similar experiment
to that performed in \cite{kato2004c} could be performed with a Si
sample. The orbital current will create two regions at the edge of
the sample where electrons occupy orbitals polarized in opposite
directions and thus will have different optical properties with
respect to circularly polarized light. The change in the angle of
the beam reflected (or transmitted, in the case of Faraday
rotation) from the surface of the sample gives information about
the orbital moment polarized on the direction of the beam. One
obstacle to performing this measurement in Si is the possibly
short relaxation time of the orbital angular momentum. Since there
are no systematic study of the orbital relaxation in Si, we take
the hole spin relaxation time as a rough estimate, since these two
quantities transform the same way, and couple to the crystal
momentum in the same way. The resolution of the Kerr microscope in
\cite{kato2004c} is $\approx 1\ \mu m$ and must be comparable to
the size of the region where the orbitally polarized electrons
accumulate. This size is $L= \sqrt{D \tau_s}$ where $\tau_s$ is
the spin-relaxation time (although this is an orbital
polarization, we expect the relaxation time to be comparable to
the spin relaxation time). $D$ is the hole diffusion coefficient
and has the expression $v_F^2 \tau/3$ where $\tau$ is the momentum
relaxation time and $v_F$ is the Fermi velocity. The size of the
orbital polarization region hence depends heavily on the hole-spin
relaxation time. Hole-spin relaxation times in semiconductors have
been measured to be anywhere from $\tau_s\approx 4\
ps$\cite{damen1991} to $\tau_s \approx 1\ ns$
\cite{roussignol1994}. Hole-spin relaxation times for Si
structures have been measured to be on the order of $\sim 10\ ps$
at low temperatures\cite{gusev1984,andrievskii2003}. However,
these measurements are ``bipolar'' measurements where both
electrons and holes are excited and spin-polarized. A
``monopolar'' spin measurement, which excites carriers only in
intraband or intrasubband transitions, would measure spin
relaxation times without electron-hole interaction and exciton
formation\cite{ganichev2002}. This measurement was carried out in
\cite{ganichev2002} for p-type quantum wells in a regime of
intraband or intersubband transitions and they measured a
hole-spin lifetime of $\sim 30\ ps$. This regime is the most
relevant for our calculations so we will use this value in our
subsequent estimations. We have estimated the hole-spin diffusion
constant (which we expect to be close to the orbital diffusion
constant) and the spatial distribution of the orbital moments. The
values for the length over which the holes are distributed are
given in Table I as $\ell$. For a steady electric current $J_y$ we
can estimate the orbital current density to be $j_I\ \sim\
(\sigma_I/\sigma_c)J_y$\cite{murakami2003}. The values of orbital
polarization density are given by the expression
$j_I\tau_s$\cite{murakami2003}. Assuming that $3\times10^{5} V/cm$
is an upper bound for the electric field in
Si\cite{levinstein1996} we have calculated several values of the
maximum orbital polarization density in Table I under $\rho_I.$

We would like to thank Prof. Z. Fang for numerous discussions on
the subject. B.A.B. acknowledges support from the Stanford
Graduate Fellowship Program. T.L.H. acknowledges support from the
NSF Graduate Research Fellowship Program. This work is supported
by the NSF under grant numbers DMR-0342832 and the US Department
of Energy, Office of Basic Energy Sciences under contract
DE-AC03-76SF00515.

%\bibliography{orbitron,extra}

\end{document}